\documentclass[prd,aps,epsf,amsfonts,floats,twocolumn,amssymb,amsmath,groupedaddress,showpacs,floatfix,nofootinbib]{revtex4}

\usepackage[]{latexsym}
\usepackage{bm}
\usepackage{subeqnarray}

\usepackage{graphicx,amssymb,amsmath}
\usepackage[english]{babel}
\usepackage{graphics}                 
\usepackage{color}                    
\usepackage{hyperref}

\newcommand{\be}{\begin{equation}}\newcommand{\ee}{\end{equation}}
\newcommand{\bea}{\begin{eqnarray}}\newcommand{\eea}{\end{eqnarray}}
\newcommand{\bsa}{\begin{subeqnarray}}
\newcommand{\esa}{\end{subeqnarray}}
\newcommand{\brr}{\begin{array}}\newcommand{\err}{\end{array}}
\newcommand{\bit}{\begin{itemize}}\newcommand{\eit}{\end{itemize}}
\newcommand{\ben}{\begin{enumerate}}\newcommand{\een}{\end{enumerate}}

\newcommand{\ba}{\begin{array}}
\newcommand{\ea}{\end{array}}

\def\lab{\label}
\def\lf{\left}

\def\non{\nonumber}
\def\rar{\rightarrow}
\def\ri{\right}\def\ti{\tilde}

\def\de{\delta}

\def\La{\Lambda}
\def\om{\omega}

\def\1{{_{1}}}\def\2{{_{2}}}

\def\noHe0{:\;\!\!\;\!\!:H_e(0):\;\!\!\;\!\!:}
\def\noHm0{:\;\!\!\;\!\!:H_\mu(0):\;\!\!\;\!\!:}

\def\lab{\label}

\def\lf{\left}

\def\non{\nonumber}

\def\rar{\rightarrow}
\def\ri{\right}\def\ti{\tilde}

\def\de{\delta}

\def\La{\Lambda}
\def\om{\omega}

\def\1{{_{1}}}\def\2{{_{2}}}

\begin{document}

\title{Probing Hawking and Unruh effects and
quantum field theory in curved space
\\ by geometric invariants.}

\author{ Antonio Capolupo${}^{\natural}$}
\author{ Giuseppe Vitiello${}^{\flat}$}
 \affiliation{${}^{\natural}$ Dipartimento di Ingegneria Industriale,
  Universit\'a di Salerno, Fisciano (SA) - 84084, Italy}
   \affiliation{${}^{\flat}$
  Dipartimento di Fisica E.R.Caianiello
  Universit\'a di Salerno, and INFN Gruppo collegato di Salerno, Fisciano (SA) - 84084, Italy}


\begin{abstract}

The presence of noncyclic geometric invariant is revealed in all the phenomena where particle generation from vacuum or vacuum condensates appear.
Aharonov--Anandan invariants  then can help to study such systems and can represent a  new tool to be used in order to provide laboratory evidence of phenomena particulary hard to be detected, such as  Hawking and Unruh effects and some features of quantum field theory in curved space simulated by some graphene morphologies. It is finally suggested that a very precise quantum thermometer can be built by exploiting geometric invariants properties.

\end{abstract}
\maketitle

\section{Introduction}

Although intensive studies have been devoted to several gravitational effects, like Unruh \cite{Unruh:1976db}, Hawking \cite{Hawking:1974sw} and Parker \cite{Parker:1968mv}  (see also \cite{Schrodinger}) effects  and to some features of quantum field theory (QFT) in curved space, these effects and features remain very hard to be directly observed or detected. They share interesting formal properties with systems described in terms of vacuum condensate and with thermal field theory \cite{Takahashi:1974zn} and also with mixing of particles \cite{Blasone:1995zc,Capolupo:2006et,Capolupo:2010ek}, dissipative systems  \cite{Celeghini:1991yv} and graphene physics \cite{Iorio:2010pv}. A common tool entering the description of all these systems is the Bogoliubov transformation  \cite{Takahashi:1974zn,BJV,MSV,Proukakis:2008}.\\
\indent An apparently separated research line is represented by the study of geometric phases and invariants \cite{Berry:1984jv}--\cite{Capolupo:2011rd} experimentally
observed in the evolution of many physical systems, from photons in optical fibers \cite{Tomita} and nuclear magnetic resonance \cite{Jones},
to superconducting circuits \cite{Leek} and electronic harmonic oscillators \cite{Pechal}.
\\
\indent In this paper we show that the non-cyclic geometric invariant is actually present in all the systems described by Bogoliubov transformations and that the Aharonov--Anandan invariant (AAI) \cite{Anandan:1990fq} could be useful in the laboratory detection of those phenomena mentioned above whose observation appears at the present state of affairs quite difficult from an experimental point of view.

We show that the Hawking effect may be revealed by means of an interferometric measurement
on a Bose-Einstein condensate. By using the experimental setup presented in \cite{Lahav}, in which an acoustic black hole has been created in a Bose-Einstein condensate accelerated by an external potential, we propose an interferometer built by two devices similar to the one in \cite{Lahav}.
In such an interferometer the accelerated Bose-Einstein condensate follows two different paths.
In one branch, the condensate is accelerated to velocities which exceed the sound velocity in order to generate a sonic event horizon and to produce Hawking radiation; in the second branch, the condensate moves at subsonic speed and then there is no event horizon. We show that the detection of a difference of geometric invariant between the two paths is related to the Hawking radiation.
In a similar way, we show that AAI due to the acceleration of two levels atom can be observed thorough interference with an inertial atom and can show the evidence of Unruh effect.

We display that AAI could be used to reveal in laboratory aspects of quantum field theory in curved space by using particular graphene morphologies.

Finally, we show that a very precise thermometer can be built by exploiting the AAI arising in the interaction of two level atoms in different thermal states.

The paper is structured as follows.
In Sec.II we recall some basic facts about AAI and Bogoliubov transformations and derive the general expression of AAI for systems represented by a Bogoliubov transformation.
In Sec.III we present the geometric invariant arising in thermal states and study its possible application to the Hawking effect in acoustic black hole. The study of AAI arising in graphene physics and in an accelerated system showing the Unruh effect are presented in Secs.IV and V, respectively. The possibility to build a very precise quantum thermometer by using AAI is analyzed in Sec.VI. Sec.VII is devoted to conclusions.

\section{Aharonov--Anandan invariant and vacuum condensates}

In order to generate the Aharonov-Anandan phase in  the course of the time evolution of an isolated system  it is necessary and sufficient that its state $|\phi(t)\rangle$ is not a stationary state, i.e. it has a nonzero value of the uncertainty $\Delta E(t) $  in energy,
$
\Delta E ^{2}(t) = \langle \phi(t)|H^{2}|\phi(t)\rangle -  (\langle \phi(t)|H|\phi(t)\rangle)^{2}.
$
 The AAI  is then defined as
$
S = ({2}/{\hbar}) \int_{0}^{  t}  \Delta E (t^{\prime}) \, dt^{\prime}\,.
$

In the phenomena mentioned in the Introduction, the  physically relevant states $|\Psi (\theta)\rangle$, with $\theta \equiv \theta (\xi, t)$ and $\xi$ some physically relevant parameter, are related to the original ones  $|\psi(t)\rangle$ by the Bogoliubov transformation $|\Psi(\theta)\rangle = J^{-1} (\theta)|\psi(t)\rangle $ (see Appendix A).
For them the variance of the energy is always different from zero. The operator $J^{-1} (\theta)$ is the  generator of the Bogoliubov transformation,
\bea\non
 \alpha^r_{\mathbf{k}} (\theta) = J^{-1} (\theta)\,a^r_{\mathbf{k}}(t) J(\theta)
 = U_{\mathbf{k}} (\theta) \, a^r_{\mathbf{k}}(t) + V_{\mathbf{k}}(\theta) \, a^{r\dagger}_{-\mathbf{k}}( t)\,.
\eea
Our discussion, however, is not limited to the single mode $a^r_{\mathbf{k}}$. It also covers distinct
mode cases ($a^r_{\mathbf{k}}$ and $b^r_{\mathbf{k}}$).
In the following we consider Bogoliubov transformations at the quantum level (not at the classical one) and we do not specify further the form of the Bogoliubov coefficients $U_{\mathbf{k}} $ and $V_{\mathbf{k}}$, except for the fact that they satisfy the relation
$|U_{\mathbf{k}} |^2 \pm |V_{\mathbf{k}} |^2=1$, with $+$ for fermions and $-$ for bosons, which guaranties that the transformation is a canonical one.
For bosons these coefficients enter the energy variance of $|\Psi(\theta)\rangle$ as the energy variance is $\Delta E(t)=\sqrt{2} \hbar\omega_{\bf k} |U_{\bf k}(\theta)| |V_{\bf k}(\theta)|$ and the geometric invariant is
\bea\label{s1}
S(t)=2\sqrt{2}\int_{0}^{  t} \omega_{\bf k} |U_{\bf k}(\theta^{\prime})| |V_{\bf k} (\theta^{\prime})| dt^{\prime}\,,
\eea
 with $\theta^{\prime} \equiv \theta (\xi, t^{\prime})$.
$S$ is related to the number of particles condensed in the vacuum  $|0(\theta)\rangle = J^{-1} (\theta)|0\rangle$,
($|0\rangle$ is the vacuum annihilated by $a_{\bf k}^r$ and $a_{-\bf k}^r$) given by
\bea
{\cal N}_{a_{\bf k}}^r(\theta)=\langle 0(\theta)|a^{r \dag}_{\mathbf{k}}a^r_{\mathbf{k}}|0(\theta)\rangle=|V_{\bf k}(\theta)|^2\,.
\eea
 Indeed
  $\Delta E(\theta)$ can be written as
\bea
\Delta E(\theta)  = \sqrt{2}\, \hbar \omega_{\bf k} \sqrt{[1 + {\cal N}_{a_{\bf k}}(\theta)] {\cal N}_{a_{\bf k}}(\theta)}\,.,
\eea
 then the corresponding AAI  is
\bea\label{s2}
S(t)=2\sqrt{2}\int_{0}^{  t} \omega_{\bf k} \sqrt{[1 + {\cal N}_{a_{\bf k}}(\theta^{\prime})] {\cal N}_{a_{\bf k}}(\theta^{\prime})}\,dt^{\prime}\,.
\eea
For fermions one obtains relations similar to Eqs.(\ref{s1})-(\ref{s2}).
In the cases of Hawking and Unruh effects, the AAI are related to  the number of particle produced in the vacuum. For these phenomena, we get a non-zero $ S$ even though the final amount of particle created are extremely small. This represents another important characteristic of the AAI.

Finally, we note that the noise does not affect AAIs for any system which presents a condensate structure.
For such systems, the background noise is given by the non-zero energy of the vacuum condensate $ |0(\theta)\rangle $.
Denoting with $H$ the Hamiltonian of the system, the noise is
\bea\label{noise}
\Lambda = \langle 0(\theta)|: H :|0(\theta)\rangle\,
\eea
 where the symbol $:...:$ denotes the normal ordering with respect to  $ |0\rangle $.
In general the noise $\Lambda$ is a c-number, (we will compute it explicitly for thermal state) an then it
does not modify the uncertainty $\Delta E(t) $ in the energy of the system and consequently it does not modify the AAI.
Indeed, it is immediate to show that the shift $H \rar H + \La$ does not affect the value of $\Delta E (\theta)$  computed on the Bogoliubov transformed state
 $|\Psi(\theta)\rangle$:
\bea\non
\Delta E^{2}_{\La} (\theta) &=& \langle \Psi(\theta)|(H + \Lambda)^{2}|\Psi(\theta)\rangle
\\ \non
&-&  (\langle \Psi(\theta)|(H + \Lambda)|\Psi(\theta)\rangle)^{2}
\\ \non
  &=& \Delta E^{2}(\theta).
  \eea
This fact leaves unchanged AAI, as can be seen by the definition of $S$. On the contrary, the noise affects Berry-like phases since $\Lambda$ contributes to the value of the Hamiltonian responsible of the system evolution.

We now analyze specific cases.

\section{Aharonov--Anandan invariant and Hawking effect}

 {\it Thermal state} -- In the formalism of Thermo Field Dynamics (TFD) \cite{Takahashi:1974zn}, (for the convenience of the reader we comment on the TFD formalism in Appendix B) the Bogoliubov  parameter  $\theta$ is related to temperature and for bosons (for fermions one can proceed in a similar way)
$U_{\bf k } = \sqrt{{e^{\beta\hbar \omega_{\bf k} }}/{(e^{\beta\hbar \omega_{\bf k } }-1)}}$, $V_{\bf{k}} = \sqrt{{1}/{(e^{\beta\hbar \omega_{\bf k} }-1})}$, with $\beta = 1/ k_{B}T$. We remark that here and in the following cases $U_{\bf k }$ and $V_{\bf k }$ are real quantities.

The uncertainty in the energy of the temperature dependent state
$
|\Psi(\theta)\rangle\, \equiv \alpha^{\dag}(\theta)|0(\theta)\rangle\,,
$
is given by
\bea\non
\Delta E(\theta) = \sqrt{2}\,\hbar \omega_{\bf k} \,U_{\bf k}(\theta)V_{\bf k}(\theta)
= \sqrt{2}\hbar \omega_{\bf k} \,\frac{{e^{ \hbar \omega_{\bf k}/2 k_{B}T }}}{{(e^{ \hbar \omega_{\bf k }/  k_{B}T }-1)}}\,.
\eea
and the AAI is
\bea
S_{T}(t) = 2\sqrt{2} \omega_{\bf k} t \,\frac{{e^{ \hbar \omega_{\bf k}/2 k_{B}T }}}{{(e^{ \hbar \omega_{\bf k }/  k_{B}T }-1)}}\,.
\eea
Then, the difference of geometric invariants
existing between two thermal states at different temperature $T_1$ and $T_2$  is
 \bea\label{DeltaphaseTermal}
\Delta S_{T}(t) = 2\sqrt{2} \omega_{\bf k} t \lf[\frac{e^{ \hbar \omega_{\bf k}/2 k_{B} T_{1} }}{e^{ \hbar \omega_{\bf k }/  k_{B}T_{1} }-1}-
\frac{e^{ \hbar \omega_{\bf k}/2 k_{B}T_{2} }}{e^{ \hbar \omega_{\bf k }/  k_{B}T_{2} }-1}\ri].
\eea
{\it Hawking effect} --
The geometric invariant difference (\ref{DeltaphaseTermal}) could help to detect the Hawking radiation in  acoustic  black hole created in Bose-Einstein condensate.\\
\indent According to the no-hair conjecture, Hawking radiation emitted by black holes  depends only on the mass, angular momentum and charge of the black hole.
The thermal bath observed outside the event horizon of a black hole has  temperature $T_{H}={\hbar c^{3}}/{(8 \pi G M k_{B})}$, where $G$ is the gravitational constant and $M$ the black hole mass.\\
\indent Recently, an acoustic black hole has been created  in a Bose-Einstein condensate of $10^{5}$ atoms of $\;^{87}$Rb  \cite{Lahav}.
In this system, the condensate is accelerated by a step-like potential to velocities which cross and exceed the speed of sound. The sonic event horizon is represented by the point where the flow velocity equals the speed of sound and the sound waves cannot escape the event horizon.
 The effective temperature of Hawking radiation is
$T_{H}=\mu / k_{B} \pi \lambda$,
where $\mu = m c^{2}$ is the chemical potential of the condensate with $m$ atomic mass, $c$ speed of sound in the condensate (typically, a few $10^{-3}m/s$ in Bose-Einstein condensate) and  $\lambda$ is related to the number of correlation lengths needed to have a thermal spectrum for the Hawking radiation. $\lambda = 7$ and $T_{H} \in (2-10) ~nK$ in \cite{Lahav}. Since the temperature of the condensate $T_{cond}$ is $(20-170) ~nK$, the Hawking radiation is very difficult to identify since it is indistinguishable from thermal noise. However, we remark that
a geometric invariant is associated to Hawking radiation. This invariant is analogous to the phase studied in the thermal state case. Now the temperature is $T\equiv T_{H}$. The geometric invariant could be detected  via interferometry  measuring the difference between geometric invariants associated to two flows of one-dimensional Bose-Einstein condensate, one in which is realized an acoustic horizon and the other in which the stream is subsonic. The realization of such configuration can be obtained by using two devices like the one  presented in \cite{Lahav}. In this way the Hawking radiation can be revealed by means of the presence of a difference of geometric invariants $\Delta S_{H}$  given by Eq.~(\ref{DeltaphaseTermal}). The relevant temperatures are now $T_{1}\equiv T_{H}+T_{cond}$ and $T_{2}\equiv T_{cond} $ since the second flow is subsonic and there is no Hawking radiation.
Fig.~1 is thus derived by plotting $\Delta S_{H}$ {\it vs} the excitation energy for different black hole temperatures.
In our computation we have taken into account that  Hawking radiation should have a wavelength shorter than the dimension of the black hole \cite{cirac} and that the minimum trapped wavelength in the experiment of ref.\cite{Lahav} is  $ \sim 1.6 \mu m$. By using Bogoliubov relation of dispersion  $\omega^{2} = c^{2}k^{2}\lf(1+{k^{2}}/{k^{2}_{c}} \ri)$, where $k _{c}={(m c)}/{\hbar}$ is the acoustic Compton wavenumber, we have derived the minimum value of the excitation energy in acoustic black hole. In Fig.~1
we considered  the fact that the black hole horizon is maintained for about $20 ~ms$ in the experiments reported in \cite{Lahav}.\\
\begin{figure}
\begin{picture}(300,180)(0,0)
\put(10,20){\resizebox{9 cm}{!}{\includegraphics{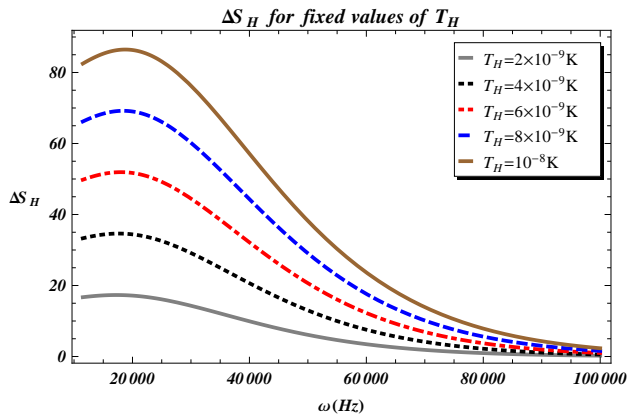}}}
\end{picture}\vspace{-1cm}
\caption{\em Plots of $\Delta S_{H}$ {\it vs} the excitation energy $\omega$, for a time interval $t = 20 ms $ and for sample values of  $T_{H} \in  [2, 10] nK$, as indicated in the inset and $T_{cond}=40nK$.}
\label{pdf}
\end{figure}
\\
\indent Fig.~1 shows that in a wide frequency interval, AAI are in principle detectable and can represent a new method to reveal Hawking effect in Bose-Einstein condensate.

We note that the finite temperature effects and deviations from Bogoliubov treatment, e.g. at
Bogoliubov-Hartree-Fock-Popov or Zaremba-Nikuni-Griffin levels, do not modify considerably
the results above presented.
Indeed, for $T \ll T_{cond}$, finite temperature effects are going to be negligible for Bose-Einstein condensate since the depletion of the Bose-Einstein condensate is negligible \cite{Stringari-Pitaevskii}.
Then, the number of condensed particle can be assumed approximatively constant in the time interval of about $20 ~ms$ \cite{Billam}. Thus, even in the non-equilibrium situations and in particular, in the case of Hawking radiation emission, the Bose occupation number can be assumed almost equal to the one obtained in the thermal equilibrium case at temperature $T_{cond}+T_{H}$,
\bea
{\cal N}_{\bf k}^r(\theta) \sim \frac{1}{(e^{\beta\hbar \omega_{\bf k} }-1)}\,.
\eea

In this case, the noise $\Lambda$ accounting for the background (original) noise and for the additional noise amplified by the horizon is obtained by computing the energy of the quanta condensed in the thermal vacuum $|0(\theta)\rangle $ at $T = T_{cond}
+ T_H \equiv 1/k_B \, \beta$, i.e.
 \bea\non
  \Lambda & = &  \sum_{r} \int d^{3}k \, \hbar \, \omega_{k}\,{\cal N}_{\bf k}^r(\theta)
 \sim \int d^{3}k \,\frac{\hbar \, \omega_{\bf k}}{e^{\beta \hbar \omega_{\bf k}}-1}\,.
  \eea
This is a c-number leaving unchanged AAI, thus AAI  provides in principle a tool able  to distinguish Hawking radiation from thermal noise.

\section{Aharonov--Anandan invariant and graphene physics}

Recently as been shown that a sheet of graphene shaped as the Beltrami pseudosphere, $dl^{2}=du^{2} + r^{2} e^{2 u / r}dv^{2}$, with $v\in [0,2\pi] $, $u\in [-\infty,0]$ and $r$ radius of curvature, displays a finite temperature electronic local density of states \cite{iorio2012} given by
\be
\rho = \frac{4}{\pi (\hbar\, v_F)^2} \frac{E \, e ^{-2u/r}}{e^{E/(k_B T_{0}\,e^{-u/r})} - 1} ~,
\ee
with $E = \hbar\, \om$. It has be shown that the electric properties can be expressed in terms of massless, neutral $(2+1)$-dim Dirac pseudoparticles, which in turn leads us to consider a general relativistic-like space-time. In particular, the electronic thermal spectrum appears to be of Hawking-Unruh type and depends on the curvature of the surface. The temperature in this case is $T = T_{0}e^{ u / r}$, with, $T_{0} = \hbar v_{F}/(2 \pi k_{B}r)$, and  $v_{F}$  Fermi velocity.
In this sense, graphene provides a realization of QFT in curved space-time.
\\
\indent
By resorting to the above results presented in detail in \cite{iorio2012}, we now proceed as in the thermal state case (with a temperature, $T = T_{0}e^{ u / r}$) and derive the AAI for electrons in graphene.
In \cite{iorio2012} an experiment with a Scanning Tunneling Microscope (STM) has been also proposed to detect the effect there  provided.
Here, by limiting ourselves to the Beltrami pseudosphere metrics, we show that such an effect could be in principle revealed also by means of AAI. Indeed, from the analysis of AAI, one can obtain the values of the temperature $T_0$.
\\
\indent The AAI
in graphene shaped as the Beltrami pseudosphere is, as in the thermal state case,
\bea\label{graphene}
S_{g}(t) = 2\sqrt{2} \omega_{\bf k} t \frac{e^{ \hbar \omega_{\bf k}/2 k_{B}T }e^{- 2 u/r }}{e^{ \hbar \omega_{\bf k }/  k_{B}T }-1}\,,
\eea
where the factor $e^{- 2 u/r }$ is due to the Weyl symmetry transformation of the $U_{\bf k}$ and $V_{\bf k}$ Bogoliubov coefficients.

Notice that  in the $r \rightarrow \infty$ limit the finite temperature electronic local density of states does not match the flat one \cite{iorio2012}. Consequently, in  such a limit Eq.~(\ref{graphene}) does not reduce to the geometric invariant of the flat space case.
\\
\indent
At given  $E$ and given $r$, the difference of invariants $\Delta S_{g}$ between  the tunneling currents along two different meridians $u$ and $u^{\prime}$ can be measured. One derives the value of  $T_{0}$ from such a measured $\Delta S_{g}$ value.
In Fig. 2 we plot $\Delta S_{g}$ vs $T_{0}$ for the values of the electron energy shown in the inset, and for $u=-r/2$ and $u^{\prime}=-r/3$, with  variable $r$.
\begin{figure}
\begin{picture}(300,180)(0,0)
\put(10,20){\resizebox{9 cm}{!}{\includegraphics{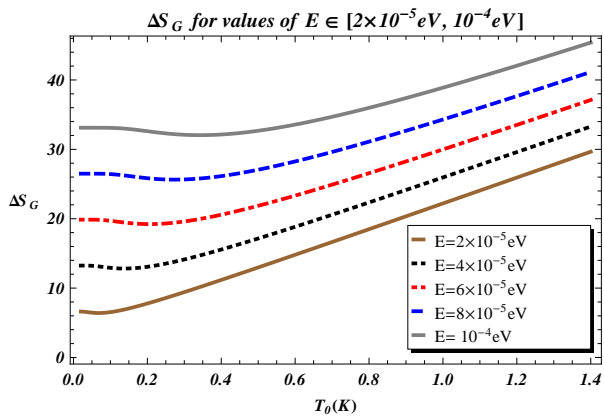}}}
\end{picture}\vspace{-1cm}
\caption{\em Plots of $\Delta S_{G}$ {\it vs} $T_{0}$, for a time interval $t = 10^{-10}s $ and for sample values of  $E \in  [2 \times 10^{-5} eV,    10^{-4} eV]$, as indicated in the inset.}
\label{pdf}
\end{figure}
We see that a non-zero difference of invariants $\Delta S_{g}$ in graphene appears for a wide range of values of $T_0$ and then it could help to verify quantum field theory in curved space in table top experiments.

\section{Aharonov--Anandan invariant and Unruh effect}

In the Unruh effect, the ground state for an inertial observer is seen by the uniformly accelerated observer as in thermodynamic equilibrium with a non-zero temperature. The Bogoliubov coefficients allow to express the Minkowski vacuum in terms of Rindler states and the temperature of thermal bath depends on the acceleration $a$ of the observer, $T_{U}={\hbar a}/{(2 \pi c k_{B}) }$.
The detection of such phenomenon is very hard: acceleration of the order of  $2.5 \times 10^{20}m/s^{2}$ corresponds to a temperature of $1 K$.
It has been shown that the Berry phase variation due to the acceleration of a two level atom, which can be observed through interference with an inertial atom, may produce evidence of the Unruh effect \cite{Ivette1,Hu-Yu}.
We observe, however, that the Berry phase is defined only for systems which have an adiabatic and cyclic evolution
and, in contrast to AAI, it is affected by thermal noise. Here we suggest that the AAI, which is independent of the particular time evolution of the system and generalize the Berry phase to the noncyclic case, may provide a  useful tool   in the laboratory detection of the Unruh effect.

By resorting to the results of \cite{Hu-Yu}, we consider a nonunitary evolution of an accelerated two level atom and we study the interaction of the atom with all vacuum modes of the electromagnetic field in the multipolar scheme.
The accelerated atom is treated as an open system in a reservoir of electromagnetic fields.
The environment induces quantum mechanical decoherence and dissipation thus implying a nonunitary evolution of the atom.

In the following, we compute the AAI of the two level system in the presence of an acceleration and in the inertial case by using the discussion presented in \cite{Hu-Yu}. The difference between the two invariant gives the geometric invariant due only to the atom acceleration.

The Hamiltonian of the system (atom plus reservoir) is
\bea
H = \frac{\hbar}{2}\,\omega_{0}\,\sigma_{3}\,+\,H_{\phi}\,-\,e \sum_{mn}\mathbf{r}_{mn}\cdot \mathbf{E}(x(t))\sigma_{mn}\,,
\eea
where $\sigma_{3}$ is the Pauli matrix, $\omega_{0}$ is the energy level spacing of the atom, $H_{\phi}$ is the Hamiltonian of the electromagnetic field, and  $\mathbf{E}$ is the electric field strength. We assume a weak interaction between atom and field, and study the evolution of the total density matrix $\rho_{tot}= \rho(0)\otimes |0\rangle \langle 0|$, in the frame of the atom.
 $\rho(0)$ is the initial reduced density matrix of the atom and $|0\rangle$ is the vacuum. The evolution, in regime of weak interaction, can be expressed in the Kossakowski-Lindbland form as,
\bea\label{evolution}
\frac{\partial \rho(\tau)}{\partial \tau} = -\frac{i}{\hbar}[H_{eff},\rho(\tau)]+ L[\rho(\tau)]
\eea
where $\tau$ is the proper time. Here, $H_{eff}$ is the effective hamiltonian,
\bea
H_{eff}\,=\,\frac{\hbar}{2}\,\Omega\,\sigma_{3}\,=\,\frac{\hbar}{2}\,\lf [\omega_{0} + \frac{i}{2} \lf(K(-\omega_{0})-K( \omega_{0})\ri) \ri] ,
\eea
where $\Omega$ is the renormalized energy level spacing which contains the Lamb shift term \cite{Hu-Yu}, $K(\omega_{0})$ is the Hilbert transform
of the correlation functions
\bea\non
G^{+}(x-y)=\frac{e^{2}}{\hbar^{2}}\sum_{i,j=1}^{3} \langle +|r_{i}|-\rangle \langle -|r_{j}|+\rangle \langle 0 |E_{i}(x) E_{j}(x)|0 \rangle\,,
\eea
i.e.
\bea
K(\omega_{0})\,=\,P \frac{1}{\pi i}\int_{-\infty}^{\infty} d \omega \frac{G(\omega)}{\omega-\omega_{0}}\,,
\eea
and $G(\omega)$ the Fourier transform of $G^{+}(x-y)$,
\bea
G(\omega)\,=\,\int_{-\infty}^{\infty} d \tau e^{i \omega \tau}G^{+}(x(\tau))\,.
\eea
In Eq.(\ref{evolution}), $L[\rho] $ is given by
\bea
L[\rho]\,=\,\frac{1}{2}\sum_{i,j=1}^{3} a_{i j } \lf(2 \sigma_{j} \,\rho\, \sigma_{i} - \sigma_{i}\, \sigma_{j}\, \rho - \rho \, \sigma_{i}\, \sigma_{j} \ri)\,,
\eea
where the coefficients of the Kossakowski matrix $a_{i j } $ are
\bea
a_{i j } \,=\, A \delta_{i j } -i B \epsilon_{i j k } \delta_{k 3} - A \delta_{i 3}\delta_{j 3}\,,
\eea
with
\bea
A = \frac{1}{4}\lf[G(\omega_{0})+ G(-\omega_{0})\ri],\;\; B = \frac{1}{4}\lf[G(\omega_{0})- G(-\omega_{0})\ri].
\eea
By expressing $\rho$ in terms of Pauli matrices as
\bea\label{ro-si}
\rho(\tau) = \frac{1}{2}\lf(1 + \sum_{i,j=1}^{3} \rho_{i}(\tau) \sigma_{i} \ri)\,,
\eea
introducing Eq.(\ref{ro-si}) in Eq.(\ref{evolution}) and assuming that the initial state of the two level atom is
 \bea
|\psi(0)\rangle = \cos \lf(\frac{\theta}{2}\ri)|+\rangle + \sin \lf(\frac{\theta}{2}\ri)|-\rangle \,,
  \eea
 the reduced density matrix can be written as
\begin{widetext}

\bea\label{matriceDensit}
\rho(\tau) =
\left(
  \begin{array}{cc}
    e^{-4 A \tau} \cos^{2} \lf(\frac{\theta}{2}\ri)+\frac{B-A}{2 A}( e^{-4 A \tau} -1) & \frac{1}{2}e^{-2 A \tau -i \Omega \tau}  \sin \theta \\
  \frac{1}{2}e^{-2 A \tau + i \Omega \tau}  \sin \theta  & 1- e^{-4 A \tau} \cos^{2} \lf(\frac{\theta}{2}\ri)-\frac{B-A}{2 A}( e^{-4 A \tau} -1)  \\
  \end{array}
\right)\,.
\eea

\end{widetext}
The eigenvalues of the matrix (\ref{matriceDensit}) are
\bea
\lambda_{\pm} = \frac{1}{2}(1 \pm \eta)\,,
\eea
where
 \bea
 \eta &=& \sqrt{\rho^{2} + e^{-4A \tau}\sin^{2}\theta}\,,
  \eea
and
 \bea
  \rho &=& e^{-4A \tau}\cos\theta + \frac{B}{A}(e^{-4A \tau}-1)\,.
 \eea
 Being $\lambda_{-}(0) = 0$, we we study the AAI only for the eigenvector corresponding to $\lambda_{+}$,
 \bea
|\phi_{+}(\tau)\rangle = \sin \lf(\frac{\theta(\tau)}{2}\ri)|+\rangle + \cos \lf(\frac{\theta(\tau)}{2}\ri)|-\rangle \,,
  \eea
  where
 \bea
 \frac{\theta(\tau)}{2} = \arctan \lf(\sqrt{\frac{ \eta + \rho }{ \eta - \rho }} \ri)\,.
 \eea
The variance of $H_{eff}$ on $|\phi_{+}(\tau)\rangle$
is given by $\Delta E(\tau) = ({1}/{2})\,\hbar\, \Omega\,\sin  \theta(\tau) $. Thus the AAI is
\bea\label{AAI}
S = \, \Omega\,\int_{0}^{t} \sin  \theta(\tau) d \tau\,.
\eea
In the case of a two-level atom uniformly accelerated in the $x$ direction with acceleration $a$,
the trajectory is
\bea
t(\tau) = \frac{c}{a} \sinh \frac{a \tau}{c}\,,\qquad x(\tau) = \frac{c^{2}}{a} \cosh \frac{a \tau}{c}\,.
\eea
The field correlation function in this case is given by \cite{Hu-Yu}
\bea
G^{+}(x,x') = \frac{e^{2}|\langle - |{\bf r}|+ \rangle|^{2}}{16 \pi^{2}\varepsilon_{0}\hbar c^{7}}\frac{a^{4}}{\sinh^{4}\lf[\frac{a}{2 c}(\tau - \tau'- i \varepsilon) \ri]}\,,
\eea
and its Fourier transform is
\bea
G(\omega) = \frac{\omega^{3} e^{2}|\langle - |{\bf r}|+ \rangle|^{2}}{6 \pi \varepsilon_{0}\hbar c^{3}}
\lf(1+\frac{a^{2}}{c^{2} \omega^{2}}\ri)\lf( 1+ \coth \frac{\pi c \omega}{a} \ri).
\eea
\begin{widetext}
Then the coefficients of the Kossakowski matrix $A_{a}$ and $B_{a}$ are \cite{Hu-Yu},
\bea
A_{a}\,=\, \frac{\gamma_{0}}{4} \,\lf(1+\frac{a^{2}}{c^{2} \omega_{0}^{2}}\ri)\,\frac{e^{2 \pi c \omega_{0}/a}+1}{ e^{2 \pi c \omega_{0}/a}-1 } \,,\qquad  B_{a}\,=\,\frac{\gamma_{0}}{4} \,\lf(1+\frac{a^{2}}{c^{2} \omega_{0}^{2}}\ri)\;,
\eea
and the AAI in Eq.(\ref{AAI}) becomes
\bea\label{faseAA}
S_{a}\,=\pm\,\int_{0}^{t}\frac{e^{2 A_{a} \tau}\sin \theta}{\sqrt{e^{4 A_{a} \tau}\sin^{2} \theta\,+\,\lf(R\,-\,R\,e^{4 A_{a} \tau}\,+\,\cos^{2} \theta \ri)^{2}}} \,\Omega_{a}\, d \tau\,,
\eea
where $R_{a} = B_{a}/A_{a}$, with
$\gamma_{0}\,=\,e^{2}\,|\langle -|\mathbf{r}|+\rangle|^{2} \omega^{3}_{0}/3\pi\varepsilon_{0}\hbar c^{3}$ the spontaneous emission rate,
and the effective level spacing of the atoms given by
\bea
\Omega_{a}\,=\,\omega_{0}\,+\,\frac{\gamma_{0}}{2\pi \omega^{3}_{0}}\,P \int_{0}^{\infty} d \omega\, \omega^{3}\lf( \frac{1}{\omega + \omega_{0}}\,-\, \frac{1}{\omega - \omega_{0}}\ri)\lf(1+\frac{a^{2}}{c^{2} \omega^{2}} \ri)\lf(1\,+\,\frac{2}{e^{2 \pi c \omega_{0}/a}-1} \ri)\,.
\eea
\end{widetext}

In the case of an inertial atom, $a=0$, the invariant $S_{a=0}$ has the same form of Eq.~(\ref{faseAA}), with $A_{a}$, $B_{a}$, $R_{a}$ replaced by $A_{0}=B_{0}=\gamma_{0}/{4}$ and  $R_{0}=1$, respectively.
The difference  between the AAIs of accelerated and inertial atoms, $\Delta S_{U} = S_{a }-S_{a=0}$, gives the geometric invariant purely in terms of the  atom acceleration.  We neglect the Lamb shift term, since the contribution of this term is very small \cite{Hu-Yu}. Moreover, we assume that the transition rate $|\langle -|\mathbf{r}|+\rangle|$ is of the order of the Bohr radius $a_{0}$ and $\omega_{0} \approx -e^{2}/8 \pi \hbar \varepsilon _{0} a_{0}$, so that $\gamma_{0}/\omega_{0} \approx 10^{-6}$ \cite{Hu-Yu}. By considering an initial state with angle $\theta = \pi /5$ and transition frequencies of the atom in the microwave regime,  $\omega_{0} \sim 10^{9} s^{-1}$, we obtain detectable phase differences  for  values of the acceleration  of order of $10^{17}-10^{18} m/s^{2}$, as shown in Fig.~3,
much less than the one required to detect Unruh radiation $(\sim 10^{26}m/s^{2})$ and  accessible with current technology.
\begin{figure}
\begin{picture}(300,180)(0,0)
\put(10,20){\resizebox{9 cm}{!}{\includegraphics{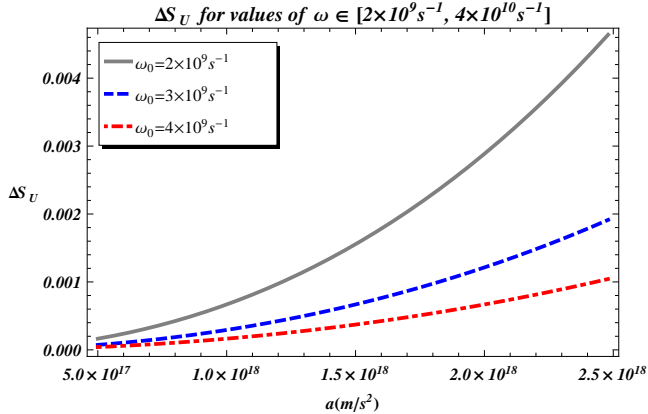}}}
\end{picture}\vspace{-1cm}
\caption{\em Plots of $\Delta S_{U}$ {\it vs} the acceleration $a$, for a time interval $t = 4 \times 2\pi /\omega_{0} $ and for sample values of  $\omega_{0} \in  [2 \times 10^{9} s^{-1},  4 \times  10^{9} s^{-1}]$, as indicated in the inset.}
\label{pdf}
\end{figure}

The above invariant differences represent Unruh effect in the interacting atomic system.
It can be detected by means of an atomic interferometer in which a two-level atom is prepared in a superposition of upper and lower states.
In one arm of the interferometer the atoms are accelerated, and in the other one they  move with constant velocity. The invariant difference is measured at the crossing point of the two arms.

Possible limitation to the detection of Unruh effect could be represented by a variation of the occupation probabilities of the atoms. However, we note that the accelerations and the time intervals here considered are enough small to have negligible probabilities of variation of  occupation \cite{Ivette1}.
\\
\indent In free field case, there is no Aharonov-Anandan phase for the  inertial system, then
the geometric invariant difference between accelerated and inertial systems, which describes the AAI in the Unruh effect, reduces to (cf. Eq.~(\ref{DeltaphaseTermal}))
$
\Delta S_{U}(t) = 2\sqrt{2} \omega_{\bf k} t \,{e^{ \pi \omega_{\bf k} c / a }}/({e^{ 2\pi \omega_{\bf k} c / a  }-1})
$.

\section{Aharonov--Anandan invariant and quantum thermometer}

We now remark that the AAI may be used in order to obtain very precise temperature measurement. We observe that the invariant (\ref{faseAA}) is acquired by the atom  also if it interacts with a thermal state.
In this case $A_{a}$ and $B_{a}$ are replaced by
\bea
A_{T}\,=\, \frac{\gamma_{0}}{4} \,\lf(\frac{1+{4\pi^{2} k_{B}^{2}T^{2}}}{{\hbar^{2} \omega_{0}^{2}}}\ri)\,\frac{e^{E_{0}/k_{B} T}+1}{e^{E_{0}/k_{B} T}-1}\,,
\eea
 and
\bea
 B_{T}\,=\,\frac{\gamma_{0}}{4} \,\lf(\frac{1+{4\pi^{2} k_{B}^{2}T^{2}}}{{\hbar^{2} \omega_{0}^{2}}}\ri)\,.
 \eea
with $E_{0} = \hbar \omega_{0}$.
Then, a thermometer can be built by means of an atomic interferometer in which a single atom follows two different paths and interacts with two thermal states (samples) at different temperatures. The difference between the  geometric invariants obtained in the two paths allows to determine the temperature of one sample once known the temperature of the other one (see also \cite{Ivette}).\\
\indent Assuming that the temperature $T_h$ of the hotter source is known, for fixed values of $\omega_{0} $ and of time, one can obtain precise measurements of the temperature $T_c$ of the colder cavity.
For the atomic transition frequencies  $\omega_{0}$ and the temperatures of the hot source reported in Fig.~4, and for time intervals of the order of
$t = 4 \times {2 \pi}/{\omega_{0}} ~s$,  measurement of cold source  temperatures can be obtained of about $ 2$ orders of magnitude below the reference temperature of the hot source, as shown in Fig.~4.\\
\begin{figure}
\begin{picture}(300,180)(0,0)
\put(10,20){\resizebox{9 cm}{!}{\includegraphics{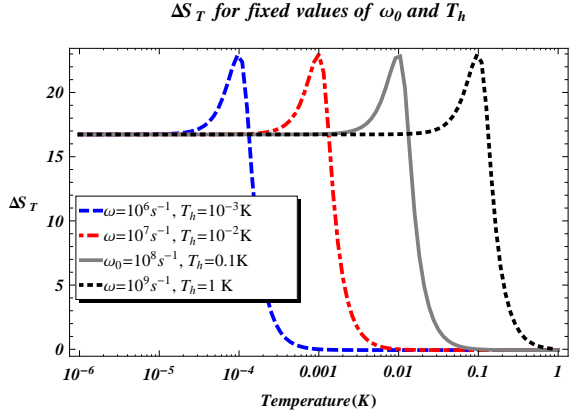}}}
\end{picture}\vspace{-1cm}
\caption{\em Plots of $\Delta S_{T}$ {\it vs} temperatures of cold sources  $T_{c}$, for sample values of  $\omega \in  [  10^{6} s^{-1}- 10^{9}  s^{-1}]$, time intervals $t = 4 \times \frac{2 \pi}{\omega} s $, and temperatures of the hot source of $T_{h} \in [ 10^{-3} K-1 K]$, as shown in the inset.}
\label{pdf}
\end{figure}

\section{Conclusions}

We have shown that all the phenomena where
vacuum condensates appear exhibit non-cyclic geometric Aharonov-Anandan invariants. We have discussed their use as novel tools in laboratory detection of phenomena particulary hard to be detected, such as  Hawking and Unruh effects, thermal field theories and graphene physics in the Beltrami pseudosphere metrics, which makes AAI also interesting in the study of QFT in curved space. Finally, we have suggested that a very precise quantum thermometer can be built by exploiting geometric invariants properties.
The AAI is a geometric invariant that can be detected in interferometric experiments when the dynamical phase is much smaller than the geometric one. Paths of slightly different lengths can be chosen in order to let the AAI be dominating over the relative dynamical phase. Present technologies allow for dynamical phase measurements within a precision of $\Delta \phi \sim 10^{-8}$, which is far smaller than the AAI values obtainable in the cases considered in this paper.

\section*{Acknowledgements}
Partial financial support from MIUR and INFN is acknowledged.

\appendix

\section{Bogoliubov transformations}

Consider a set of bosonic ladder operators $a_{\mathbf{k}}$ and $b_{\mathbf{k}}$. The canonical commutation relations (CCRs) are:
$[a_{\mathbf{k}}, a^{\dagger}_{\mathbf{p}}]=[b_{\mathbf{k}}, b^{\dagger}_{\mathbf{p}}]=\delta^{3}(\mathbf{k}-\mathbf{p})\,$,
with all other commutators vanishing. The vacuum $|0\rangle$ is defined by $a_{\mathbf{k}}|0\rangle = b_{\mathbf{k}}|0\rangle = 0$, and the Fock space is built out of it in the well known way. Such a space provides an irreducible representation of the algebra of the CCRs.

The Bogoliubov transformations on the quantum operators $a_{\mathbf{k}}$ and $b_{\mathbf{k}}$ with transformation parameter  $\theta_{\mathbf{k}}$ have the form:
\bsa\label{Atilde}
\alpha_{\mathbf{k}}(\theta) &=& U_{\mathbf{k}}(\theta) \, a_{\mathbf{k}} - V_{\mathbf{k}}(\theta) \,  b^{\dagger}_{\mathbf{k}};
\\
\beta_{\mathbf{k}}(\theta) &=& U_{\mathbf{k}}(\theta) \, b_{\mathbf{k}} - V_{\mathbf{k}}(\theta) \,  a^{\dagger}_{\mathbf{k}}.
\esa
The requirement that they leave the CCRs invariant, i.e. are canonical transformations, implies that $|U_{\mathbf{k}} |^2 - |V_{\mathbf{k}} |^2=1$, and then, $U_{\mathbf{k}}=e^{i\phi_{\mathbf{k}}}\, \cosh \theta_{\mathbf{k}}$ and $V_{\mathbf{k}}=e^{i\phi_{\mathbf{k}}}\, \sinh\theta_{\mathbf{k}}$. For simplicity we put $\phi_{\mathbf{k}} = 0$.
The transformations (\ref{Atilde}) can be rewritten  as
$
\alpha_{\mathbf{k}}(\theta) = J^{-1}(\theta)\,  a_{\mathbf{k}}\,  J(\theta)\,$, and $\beta_{\mathbf{k}}(\theta) = J^{-1}(\theta)\,  b_{\mathbf{k}}\,  J(\theta)\,$, with the generator
$J(\theta) = \exp \lf[-\sum_{\mathbf{k}} \theta_{\mathbf{k}}\lf(a_{\mathbf{k}}^{\dagger} b_{\mathbf{k}}^{\dagger} - a_{\mathbf{k}} b_{\mathbf{k}} \ri)\ri]\,$.\\
\indent The vacuum $|0(\theta)\rangle$ relative to the transformed operators $\alpha_{\mathbf{k}}(\theta)$ and $\beta_{\mathbf{k}}(\theta)$ is defined by $\alpha_{\mathbf{k}}(\theta)|0(\theta)\rangle = \beta_{\mathbf{k}}(\theta)|0(\theta)\rangle=0$ and is related to the vacuum $|0\rangle$ by
$|0(\theta)\rangle = J^{-1} (\theta)|0\rangle\,$.
The  Stone--von Neumann theorem in Quantum Mechanics guaranties that for a finite number of degrees of freedom, i.e. for discrete $\mathbf{k}$, all the representations of the CCRs are unitarily equivalent representations, i.e. in such a case $J^{-1} (\theta)$ is a unitary operator. For discrete $\mathbf{k}$ the vacua $|0\rangle$ and $|0(\theta)\rangle$ and the corresponding Fock spaces on them constructed are unitary equivalent vacua and spaces. Thus any vector in one space can be expressed in terms of  vectors in the other space.  On the other hand, in the $\mathbf{k}$ continuum limit, the Stone--von Neumann theorem does not hold and the transformation $|0(\theta)\rangle = J^{-1} (\theta)|0\rangle\,$ is only a formal relation since $|0\rangle$ and $|0(\theta)\rangle$ (and the corresponding Fock spaces) are unitarily inequivalent states (and spaces). An explicit example is provided by the thermo field dynamics (TFD) formalism briefly summarized below.
Thus, in the  $\mathbf{k}$ continuum limit,  Bogoliubov transformations  relate sets of operators whose corresponding vacuum states belong to different (inequivalent) representations of the CCRs and provide an ``exact'' formalism, different from and not to be confused with the Bogoliubov approximation formalism (discussed in full detail in ref.\cite{NJP2008}), of which the conclusions presented in the main text of the paper are independent.

\section{Thermo Field Dynamics}

In TFD  the statistical average $\langle{\cal A}\rangle$ of
an observable ${\cal A}$ is expressed as the expectation value in the temperature
dependent state $|0(\theta)\rangle$ \cite{Takahashi:1974zn}:
\bea \langle{\cal A}\rangle~ \equiv~ \frac{{\rm Tr}[{\cal A}~ e^{-{\beta}{\cal H}}]}{{\rm Tr}[e^{-{\beta}{\cal H}}]}~ =~ \langle 0(\theta)|{\cal
A}|0(\theta)\rangle \; , \label{A1} \eea
with ${\cal H}= H - \mu N$,  $\mu$ is the chemical potential, $\beta = 1/k_B T$ and $\theta \equiv \theta(\beta)$ is the transformation parameter related to temperature. For simplicity, $\mu$ will be neglected in the following. The possibility to construct such a  state $|0(\theta)\rangle$ satisfying Eq.~(\ref{A1}) is guaranteed by the very same structure of QFT allowing multi-vacua states, i.e. infinitely many unitarily inequivalent representations of the CCRs. In TFD different representations are labeled by different temperature values. One can show that Eq.~(\ref{A1}) is obtained by doubling the operator algebra ${\cal A} \rightarrow {\cal A} \otimes {\cal A}$ and, correspondingly, doubling the state space. Here, as customary in the TFD formalism, we denote the doubled states by using the tilde notation $|n\rangle \rightarrow |n\rangle \otimes |{\tilde{n}}\rangle \equiv |n, {\tilde{n}}\rangle$, where $|n\rangle$ and $|{\tilde{n}}\rangle$ are eigenstates of the Hamiltonian ${\cal H}$ and ${\tilde {\cal H}}$, respectively: ${\cal H}|n\rangle =E_{n}|n\rangle$,  $\langle n|m\rangle = \de_{nm}$, and ${\tilde {\cal H}}|\tilde n\rangle =E_{n}|\tilde
n\rangle$, $\langle \tilde n|\tilde m\rangle = \de_{nm}$.
${\tilde {\cal H}}$ has the same operatorial form of ${\cal H}$. Non-tilde and tilde operators
are assumed to be  commuting boson  operators. The state $|0(\theta)\rangle$ is found to be
\bea \lab{A2} |0(\theta)\rangle = \frac{1}{\sqrt{{\rm Tr}[e^{-{\beta}{\cal H}}]}} \sum_{n} e^{-{{\beta}
E_{n}}\over 2}|n,\tilde n\rangle ~ . \eea

Let us  consider the example of the number operator $N = a^{\dag}a$ for the boson case. For notational simplicity we neglect the momentum index. We introduce the tilde operators and the commutation relations are $[ a,a^{\dagger} ] = 1$, $[ {\tilde a} , {\tilde
a}^{\dagger} ] = 1$.
All other commutators are  zero and $a$ and ${\tilde
a}$ commute among themselves. The state $|0(\theta)\rangle$ is formally
given (at finite volume) by
\bea
\lab{A4} |0(\theta)\rangle = J^{-1}(\theta)|0\rangle  = {1 \over{U(\theta)}} \exp \Big(
\frac{V(\theta)}{U(\theta)} \Big)a^{\dag}{\tilde a}^{\dag} |0\rangle
~, \eea
where $|0\rangle \equiv |0,{\tilde 0}\rangle$ is the vacuum for $a$ and $\tilde a$,
$J^{-1}(\theta) =e^{i\theta {\cal
G}}$, with $ {\cal G} \equiv -i (a^{\dag}{\tilde a}^{\dag} - a{\tilde a})\,,$
and
$U(\theta) \equiv  \sqrt{1+f_{B}(\om) }$, $~V(\theta)
\equiv \sqrt{f_{B}(\om) }$, i.e. $U^{2}(\theta) - V^{2}(\theta) = 1$,
so that
\bea U(\theta) = \cosh \theta  ~, ~~~\quad V(\theta) = \sinh
\theta  ~ . \lab{A6} \eea
In these relations  $f_{B}(\om)$, with $\om$ the  energy of the
quantum $a$, denotes the Bose-Einstein distribution. The expectation value of $N$ in $|0(\theta)\rangle$ gives then its statistical average (cf. Eq.~(\ref{A1}) with ${\cal H} = H = \om \,a^\dag a$):
\bea \langle N \rangle~
=~ \langle 0(\theta)| N|0(\theta)\rangle = \frac{1}{e^{\beta
\om} - 1} = f_{B}(\om) \; . \lab{A3} \eea
${\cal G}$ is the generator of the Bogoliubov transformations
\bsa
\alpha(\theta) &=& e^{i{\theta}{\cal G}}\, a\,
e^{-i{\theta}{\cal G}} =  a \,{\rm cosh} ~\theta - {\tilde
a}^{\dagger}
\, {\rm sinh} ~\theta,
\\
{\tilde \alpha}(\theta) &=& e^{i{\theta}{\cal G}} \,{\tilde a}\,
e^{-i{\theta}{\cal G}}  = {\tilde a} \,{\rm cosh} ~\theta -
 a^{\dagger} \,{\rm sinh} ~\theta\,.
\lab{A7} \esa
 The commutation relations are $[ \alpha(\theta) ,
\alpha^{\dagger}(\theta) ] = [ {\tilde \alpha}(\theta) , {\tilde
\alpha}^{\dagger}(\theta) ] = 1 $. All other commutators are vanishing. $\alpha(\theta)$ and ${\tilde \alpha}(\theta)$ commute among themselves and annihilate the state $|0(\theta)\rangle$:
\bea \alpha(\theta)|0(\theta)\rangle = 0~,  ~~\qquad  {\ti
\alpha}(\theta)|0(\theta)\rangle = 0~. \lab{p324a} \eea
$|0(\theta)\rangle$ is  called the thermal vacuum and is the zero energy state for the Hamiltonian $\hat{H} \equiv H - {\ti H} = \om (a^{\dag}a - {\tilde
a}^{\dag}{\tilde a})$, i.e. $\hat{H}|0(\theta)\rangle = 0$.

When the momentum ${\bf k}$ index is restored
the operator ${\cal G}$ and the thermal vacuum are formally (at finite
volume) given by
\bea \lab{(2.11)} {\cal G} =- i \sum_{\bf k}{ ( a_{\bf
k}^{\dagger} {\ti a}_{\bf k}^{\dagger} - a_{\bf k} {\ti a}_{\bf
k}) }, \eea
\bea\lab{(2.12)} |0(\theta)\rangle = \prod_{\bf k}
{1\over{\cosh{\theta_k}}} \exp{ \left ( \tanh {\theta_k} ~a_{\bf
k}^{\dagger} {\ti a}_{\bf k}^{\dagger} \right )} |0\rangle ,
\eea
with $\langle 0(\theta) | 0(\theta)\rangle = 1, ~ \forall \theta $~.
It can be shown that $|0(\theta)\rangle$ is an $SU(1,1)$
generalized coherent state \cite{Perelomov:1986tf}.
In the infinite volume limit, the continuous limit relation $ \sum_{\bf k} \rar
{V\over{(2 \pi)^{3}}} \int \! d^{3}{k}$ gives
\bea\lab{(2.13)}
{\langle 0(\theta) | 0\rangle \rightarrow 0~\quad
{\rm as}~ V\rightarrow \infty } ~\quad \forall~  \theta =
\{\theta_k\} \neq 0 ,
\eea
for $\int \! d^{3} k ~ \ln \cosh \theta_k$ finite and
positive. In general, $ {\langle 0(\theta) | 0(\theta') \rangle
\rightarrow 0~ {\rm as}~ V\rightarrow \infty} ~~\forall~\theta$~
and~ $\theta'$, with $ \theta' \neq \theta$. The meaning of these relations is that for each  ${\theta} \equiv
\{\theta_k\}$ the representation $\{ |0(\theta)\rangle \}$ of the CCRs is  unitarily inequivalent to the
representations $\{ |0(\theta')\rangle,~\forall \theta'\neq
\theta \}$ in the infinite volume limit.
The state $|0 (\theta) \rangle $ is a condensate of pairs of  $a$ and $\ti
a$ quanta and
\bea \lab{(2.13a)}
\hspace{-4mm}{\cal N}_{a_{\bf k}}(\theta) = \langle 0(\theta)
|a_{\bf k}^{\dagger}a_{\bf k}| 0(\theta)\rangle = |V_{\bf k}(\theta)|^{2} =
 \sinh^{2} \theta_k. \eea
Similar expression is obtained for ${\cal N}_{{\ti a}_{\bf k}}(\theta)$.\\
\indent A copy $\{ \alpha_{\bf k}(\theta) , \alpha_{\bf
k}^{\dagger}(\theta) , {\ti \alpha}_{\bf k}(\theta) , {\ti \alpha}_{\bf
k}^{\dagger}(\theta) \, ; \, | 0(\theta) \rangle\, |\, \forall {k}
\}$ thus exists for each $\theta$  of the original algebra induced by the Bogoliubov generator. It generates the
group of automorphisms of ${\bigoplus_{\bf k} su(1,1)_{\bf k}}$ parameterized by $\theta_k$, namely a realization of the operator algebra at each $\theta$, which can be implemented by Gel'fand-Naimark-Segal
construction in the C*-algebra formalism \cite{Takahashi:1974zn,Bratteli:1979a}.\\
\indent In non-equilibrium systems or dissipative systems, for time dependent Bogoliubov parameter $\theta (t)$, since the $\ti a$ particle can be thought of as the ``hole" (the anti-particle) of the $a$ particle, the energy may flow out of the $a$-system  into the $\ti a$-system, or vice-versa  \cite{Celeghini:1991yv}. The tilde system thus represents the thermal bath or the environment into which the $a$-system is embedded.
\\
\indent The TFD formalism above presented may be extended to the fermion case \cite{Takahashi:1974zn}, which here we do not present for brevity.

\end{document}